\documentclass[floats,floatfix,showpacs,preprintnumbers,amssymb,prl,twocolumn,superscriptaddress,nolongbibliography,reprint]{revtex4-2}

\usepackage{amssymb,amsmath,verbatim,mathtools,needspace,enumitem,etoolbox,graphicx,physics,microtype,afterpage,xspace,tabularx,lmodern,multirow}
\usepackage{gensymb}
\usepackage[normalem]{ulem}
\usepackage[dvipsnames, usenames]{xcolor}
\usepackage{xr-hyper}
\definecolor{linkcolor}{rgb}{0.0,0.3,0.5}
\usepackage[unicode, colorlinks=true, linkcolor=linkcolor, citecolor=linkcolor, filecolor=linkcolor, urlcolor=linkcolor, linktocpage, breaklinks]{hyperref}
\usepackage[all]{hypcap}
\usepackage[T1]{fontenc}
\usepackage[utf8]{inputenc}
\usepackage[export]{adjustbox}
\usepackage[usenames,dvipsnames]{xcolor}
\hypersetup{colorlinks=true,citecolor=romared,linkcolor=romared,urlcolor=romared}

\usepackage{booktabs}

\definecolor{romared}{RGB}{142,0,28}

\newcommand{\be}{\begin{equation}}
\newcommand{\ee}{\end{equation}}

\def\be{\begin{equation}}
\def\ee{\end{equation}}
\newcommand{\beq}{\begin{eqnarray}}
\newcommand{\eeq}{\end{eqnarray}}

\usepackage{makecell}
\usepackage{soul}

\newcolumntype{Y}{>{\centering\arraybackslash}X}

\definecolor{romared}{RGB}{142,0,28}

\newcommand{\Mh}{M_{\rm halo}}
\newcommand{\MBH}{M_{\rm BH}}

\makeatletter
\newcommand*{\addFileDependency}[1]{% argument=file name and extension
  \typeout{(#1)}
  \@addtofilelist{#1}
  \IfFileExists{#1}{}{\typeout{No file #1.}}
}
\makeatother

\begin{document}
\title{Spinning black holes in astrophysical environments}
\author{Pedro G. S. Fernandes}
\affiliation{Institut f\"ur Theoretische Physik, Universit\"at Heidelberg, Philosophenweg 12, 69120 Heidelberg, Germany}
\author{Vitor Cardoso} 
\affiliation{Center of Gravity, Niels Bohr Institute, Blegdamsvej 17, 2100 Copenhagen, Denmark}
\affiliation{CENTRA, Departamento de F\'{\i}sica, Instituto Superior T\'ecnico -- IST, Universidade de Lisboa -- UL, Avenida Rovisco Pais 1, 1049-001 Lisboa, Portugal}

\date{\today}

\begin{abstract}
We present stationary and axially-symmetric black hole solutions to the Einstein field equations sourced by an anisotropic fluid, describing rotating black holes embedded in astrophysical environments. We compute their physical properties, including quantities associated with the circular geodesics of massless and massive particles, analyze their shadows and image features, and energy conditions. Overall, we find that deviations from the Kerr metric grow with spin.
\end{abstract}

\maketitle

%\tableofcontents

%%%%%%%%%%%%%%%%%%%%%%%%%%%%%%%%%%%%%%%%%%%%
\noindent \textbf{\textit{Introduction.}}
%%%%%%%%%%%%%%%%%%%%%%%%%%%%%%%%%%%%%%%%%%%%
Vacuum black hole (BH) solutions are idealized models. In reality, astrophysical compact objects exist within complex matter-rich environments, including accretion disks and dark matter halos \cite{Sadeghian:2013laa,Gondolo:1999ef}. Understanding how these environments influence a BH, affect observational signatures, and offer insights into the nature of the surrounding matter is a growing field of study.
To investigate how the astrophysical environment influences the BH geometry, one must solve the Einstein field equations with a stress-energy tensor that captures the environmental effects. In recent years, the Einstein cluster \cite{Einstein:1939ms,hoganReconstructionMinkowskianSpacetime1978,1968ApJ...153L.163Z,1974RSPSA.337..529F,Comer:1993rx,kumardattaNonstaticSphericallySymmetric1970,bondiDattasSphericallySymmetric1971,Gair:2001qu,Szybka:2018hoe,Mahajan:2007vw,Magli:1997qf} has been used as a model for dark matter halos~\cite{Acharyya:2023rnq,Boehmer:2007az,Lake:2006pp,Geralico:2012jt,Jusufi:2022jxu,Fernandes:2025lon}, and to study their impact on BH geometries~\cite{Cardoso:2021wlq,Figueiredo:2023gas,Speeney:2024mas,Cardoso:2022whc,Jusufi:2022jxu,Shen:2023erj,Shen:2024qbb,Pezzella:2024tkf,Maeda:2024tsg,Spieksma:2024voy,Konoplya:2022hbl,Macedo:2024qky,Xavier:2023exm,Dai:2023cft,Konoplya:2021ube,Ovgun:2025bol,Myung:2024tkz,Gliorio:2025cbh,Fernandes:2025lon,Joshi:2011zm}. The Einstein cluster models a large collection of non-interacting particles moving along circular geodesics in all directions under their collective gravitational field. The emergent system is described by an anisotropic fluid with zero radial pressure.

Most existing studies have been limited to spherical symmetry~\footnote{Ref.~\cite{Zhang:2024hjr} has recently attempted to obtain rotating astrophysical BH solutions. While the effort is timely and relevant, there are aspects of their approach that merit further scrutiny. In particular, the claim that the Bianchi identities lead to algebraic constraint equations for all transverse pressures does not hold, which affects the validity of the whole method. Additionally, the energy density profile is not discussed in detail, and the parameters introduced may lack clear physical interpretation, especially given the approximate coordinate transformation employed. Finally, the work offers very limited discussion on the physical properties of the solutions and provides little information on the convergence and robustness of the numerical implementation.
}, yet astrophysical BHs are expected to have non-negligible spin and to be immersed in galaxies carrying intrinsic angular momentum. Extending these models to include rotation is therefore essential to fully understand the impact of environments on astrophysical BHs. The generalization to rotating BH spacetimes poses significant challenges.
The absence of spherical symmetry seems to preclude the finding of closed form solutions. Slowly rotating anisotropic fluids are known~\cite{Cardoso:2007az,Beltracchi:2024dfb,Becerra:2024xff}, but their embedding in a BH spacetime is less trivial. Moreover, standard procedures such as Newman–Janis-inspired algorithms~\cite{Newman:1965tw,Newman:1965my,Gurses:1975vu,Beltracchi:2021ris,Beltracchi:2021tcx,Kocherlakota:2024sxx,Kim:2019hfp} generally fail to yield physically meaningful solutions, as they typically introduce pathologies or obscure the interpretation of the matter content \cite{Hansen:2013owa}. Since analytic solutions are unobtainable, numerical techniques are necessary. These, however, require solving numerically a highly non-linear system of coupled partial differential equations.
The goal of this letter is to extend the Einstein cluster to include a rotating BH, solve the field equations, and then study the properties of the resulting solutions. We work in units $c=G=1$.

%%%%%%%%%%%%%%%%%%%%%%%%%%%%%%%%%%%%%%%%%%%%
\noindent \textbf{\textit{The framework.}}
%%%%%%%%%%%%%%%%%%%%%%%%%%%%%%%%%%%%%%%%%%%%
We start by considering solutions to the Einstein field equations, $G_{\mu \nu} = 8\pi T_{\mu \nu}$, sourced by an anisotropic fluid, with stress-energy tensor
\begin{equation}
    \begin{aligned}
        T_{\mu \nu} =& \left( \varepsilon + p_{1} \right) u_\mu u_\nu + p_{1} g_{\mu \nu} \\&+ (p_r-p_{1}) k_\mu k_\nu + (p_{2} - p_{1})s_\mu s_\nu,
    \end{aligned}
    \label{eq:T}
\end{equation}
where $\varepsilon$, $p_r$, $p_1$ and $p_2$ are the energy density, radial pressure and transverse pressures in the comoving frame of the fluid, respectively. The vectors $u^\mu$, $k^\mu$ and $s^\mu$ are, respectively, the four-velocity of the fluid and spacelike vectors that define the directions of anisotropy such that $u^\mu u_\mu = -1$, $k^\mu k_\mu = 1$, $s^\mu s_\mu =1$, $u^\mu k_\mu = 0$, $u^\mu s_\mu = 0$, and $k^\mu s_\mu = 0$.
The eigenvalues of the stress-energy tensor are $-\varepsilon$, $p_r$, $p_1$ and $p_2$. This ansatz guarantees that there are no energy and momentum fluxes in the rest frame of the fluid. Such a fluid, modeling a dark matter halo in our case, is expected to generically be present near black holes as a result of accretion. As noted in Refs. \cite{Sadeghian:2013laa,Gondolo:1999ef}, the density profile produced by the (adiabatic) accretion-driven growth of a black hole typically develops an overdense cusp with a sharp cutoff in the vicinity of the horizon.

To obtain rotating BH solutions, we consider a line-element that is stationary, axially-symmetric and circular, belonging to the Weyl-Lewis-Papapetrou class. We use quasi-isotropic coordinates and the metric is described by four functions, $f$, $g$, $h$, $\omega$, of $r$ and $\theta$
\begin{equation}
    \begin{aligned}
        \dd s^2 = -f \frac{N_-^2}{N_+^2}\dd t^2  + & \frac{g}{f}N_+^4 \bigg[ h\left( \dd r^2 + r^2 \dd \theta^2 \right)\\&
        +r^2 \sin^2\theta \left( \dd \varphi - \omega \dd t \right)^2\bigg],
    \end{aligned}
    \label{eq:metric}
\end{equation}
where $N_\pm = (1\pm r_H/r)$, and $r_H$ is the coordinate location of the event horizon. 
In these coordinates, the four-velocity of the fluid can be expressed as
\begin{equation}
    u^\mu = \frac{1}{\sqrt{-(g_{tt} + 2 \Omega g_{t\varphi} + \Omega^2 g_{\varphi \varphi})}} (1,0,0,\Omega),
\end{equation}
where $\Omega\equiv \Omega(r,\theta)$ is the angular velocity of the fluid and the spacelike vectors are chosen as
\begin{equation}
    %k^\mu = \left(  0, g_{rr}^{-1/2}, 0, 0 \right), \quad s^\mu = (0,0,0,g_{\varphi \varphi}^{-1/2}),
    \begin{aligned}
        &k^\mu = \left(  0, g_{rr}^{-1/2}, 0, 0 \right),\\&
        s^\mu = \frac{1}{\sqrt{g_{\varphi \varphi} + 2 \zeta g_{t\varphi} + \zeta^2 g_{tt}}}(\zeta,0,0,1),
    \end{aligned}
\end{equation}
where $\zeta = - (g_{t\varphi} + \Omega g_{\varphi \varphi})/(g_{tt} + \Omega g_{t\varphi})$, such that the orthogonality relations hold. Note that the four-velocity of the fluid is proportional to the Lorentz factor $(1-v^2)^{-1/2}$, where $v$ is the 3-velocity of the fluid with respect to a local zero angular momentum observer (ZAMO) \cite{Paschalidis:2016vmz,Gourgoulhon:2010ju}, $v^2 = - g^{tt} g_{\varphi \varphi} \left( \Omega - \omega \right)^2$. The four-velocity of the fluid is real only if the angular velocity of the fluid, $\Omega$, is bounded between two values, $\Omega_\pm = (-g_{t\varphi}\pm \sqrt{\Delta})/g_{\varphi \varphi}$, where $\Delta \equiv g_{t\varphi}^2 - g_{tt}g_{\varphi \varphi}$, vanishes at the BH horizon \cite{Kocherlakota:2024sxx}. Therefore, at the horizon, the angular velocity of the fluid must obey a synchronization condition $\Omega = \Omega_H$, where $\Omega_H = -g_{t\varphi}/g_{\varphi \varphi}|_\mathcal{H}$, is the angular velocity of the horizon.

The metric \eqref{eq:metric} has two Killing vector fields associated with stationarity and axial-symmetry, $\eta^\mu = (\partial_t)^\mu$ and $\chi^\mu = (\partial_\varphi)^\mu$. The linear combination $\xi^\mu = \eta^\mu + \Omega_H \chi^\mu$, is orthogonal to and null on the event horizon. For a spacetime with these symmetries, the total mass of the spacetime obeys the Smarr relation \cite{PhysRevLett.30.71,Bardeen:1973gs}
\begin{equation}
    M = \frac{\kappa}{4\pi} A_H + 2\Omega_H J - 2\int d^3x \sqrt{-g} \left( T_{t}^{\phantom{t}t} - \frac{1}{2} T  \right),
    \label{eq:smarr1}
\end{equation}
where $\kappa$ is the surface gravity on the event horizon, $A_H$ the area of the event horizon, $T$ the trace of the stress-energy tensor, and $J$ is the total angular momentum given by $J = J_H + \int d^3x \sqrt{-g} T^{t}_{\phantom{t}\varphi}$, where $J_H$ is the angular momentum of the BH, computed as a Komar integral. We adopt a model for a dark matter halo similar to the Einstein cluster, where the dark matter halo has zero angular momentum. Consequently, the total angular momentum of the system is $J = J_H$. This occurs when $T^{t}_{\phantom{t}\varphi} = 0$, which is true when we choose the rotation law $\Omega = \omega$, i.e., the angular velocity of the dark matter halo is the frame-dragging angular velocity and therefore it has zero velocity with respect to the ZAMO. Note that the synchronization condition at the horizon is respected in this case.
Under these conditions, Eq. \eqref{eq:smarr1} can be rewritten as $M = M_{\rm BH} + M_{\rm halo}$, where $M_{\rm BH} = \frac{\kappa}{4\pi} A_H + 2\Omega_H J_H$, is the mass of the BH, and $M_{\rm halo}$ corresponds to the integral in Eq. \eqref{eq:smarr1}.
The total mass $M$ and angular momentum $J$ can also be computed from the asymptotic decay of the metric components $g_{tt} = -1 + \frac{2M}{r} + \mathcal{O}(r^{-2})$, and $g_{t\varphi} = - \frac{2J}{r} \sin^2\theta + \mathcal{O}(r^{-2})$, while the surface gravity and the area of the event horizon for the metric \eqref{eq:metric} are computed as
\begin{equation}
    \kappa = \frac{f}{8r_H \sqrt{g h}}, \quad A_H = 32\pi r_H^2 \int_{0}^\pi \frac{g\sqrt{h}}{f} \sin \theta \dd \theta,
    \label{eq:kappaAH}
\end{equation}
where these expressions are to be evaluated at $r=r_H$.

%%%%%%%%%%%%%%%%%%%%%%%%%%%%%%%%%%%%%%%%%%%%
\noindent \textbf{\textit{Solving the field equations.}}
%%%%%%%%%%%%%%%%%%%%%%%%%%%%%%%%%%%%%%%%%%%%
To solve the field equations and obtain stationary, axially-symmetric BH solutions, we follow the method of Ref. \cite{Fernandes:2022gde}. There are six non-trivial components of the Einstein field equations, namely the $tt$, $rr$, $\theta \theta$, $\varphi \varphi$, $t\varphi$ and $r \theta$ components. These are split into two groups. Defining $E_{\mu \nu} = G_{\mu \nu} - 8\pi T_{\mu \nu}$, we consider the following combinations of field equations: $E_{\mu}^{\phantom{\mu}\mu} - 2 E_{r}^{\phantom{r}r} - 2\omega E_{\varphi}^{\phantom{\varphi}t} = 0$, $E_{\varphi}^{\phantom{\varphi}t} = 0$, $E_{r}^{\phantom{r}r} + E_{\theta}^{\phantom{\theta}\theta} = 0$, and $E_{\varphi}^{\phantom{\varphi}\varphi} - \omega E_{\varphi}^{\phantom{\varphi}t} - E_{r}^{\phantom{r}r} -E_{\theta}^{\phantom{\theta}\theta} = 0$. These combinations ensure that each equation contains second-derivatives of only one of the four metric functions. The remaining linearly independent equations are taken to be $E_{r}^{\phantom{r}\theta}=0$, and $E_{r}^{\phantom{r}r}-E_{\theta}^{\phantom{\theta}\theta}=0$. However, as explained in Refs. \cite{Wiseman:2002zc, Herdeiro:2015gia}, these are directly related to the conservation of the stress-energy tensor, $\nabla_\mu T^{\mu \nu} = 0$, via the Bianchi identities, and therefore can be substituted by the conditions $\nabla_\mu T^{\mu r} = 0$, and $\nabla_\mu T^{\mu \theta} = 0$.
The system contains eight undetermined functions of $r$ and $\theta$, namely $f$, $g$, $h$, $\omega$, $\varepsilon$, $p_r$, $p_1$ and $p_2$. The four combinations of Einstein equations, together with the conservation of the stress-energy tensor, determine six of these functions, that we choose to be the four metric functions and the two tangential pressures $p_1$ and $p_2$ \footnote{In the spherically symmetric limit, the field equations impose $p_1=p_2$}. To close the system we choose a profile for the energy density $\varepsilon$ of the dark matter halo, and provide an equation of state relating $p_r$ and $\varepsilon$. For the equation of state we choose $p_r = 0$, in accordance with the Einstein cluster. Motivated by the Hernquist profile \cite{1990ApJ...356..359H}, and by Ref. \cite{Cardoso:2021wlq}, we use the following profile for the energy density
\begin{equation}
    \varepsilon = \frac{\mathcal{M} \left( a_0 + r_H \right)}{2\pi r (r+a_0)^3 b^5} \left(1 - \frac{r_H}{r} \right)^2,
    \label{eq:edensity}
\end{equation}
where $a_0$ is a typical length scale associated with the halo, and $\mathcal{M}$ represents the mass of the halo $M_{\rm halo}$ to leading order in $\mathcal{M}/a_0$ in the static limit. The function $b\equiv b(r)$ is a complicated function of $r$ presented in the \hyperref[suppmaterial]{Supplemental Material}, and as explained there, it guarantees that the parameters $\mathcal{M}$ and $a_0$ have suitable physical meaning. Importantly, $b>0$, and thus $\varepsilon \geq 0$ for physically relevant choices of the parameters $\mathcal{M}$ and $a_0$. For large $r$, as well as in the absence of a black hole, this profile for the energy density agrees with the Hernquist profile. This energy density profile therefore models a dark matter halo co-existing with a black hole.

The functions obey the following boundary conditions. At the horizon ($r=r_H$), by solving the field equations order-by-order in a near-horizon expansion, we find $\partial_r f = \partial_r g = \partial_r h = \partial_r p_1 = \partial_r p_2 = 0, \quad \omega = \Omega_H$, while asymptotically they obey $f=g=h=1, \quad \omega = p_1 = p_2 =0$.
Regularity, axial symmetry and parity considerations imply $\partial_\theta f = \partial_\theta g = \partial_\theta h = \partial_\theta \omega = \partial_\theta p_1 = \partial_\theta p_2 = 0$, at $\theta=0$ and $\theta=\pi/2$ \footnote{Incidentally, the field equations impose that $p_1=p_2$, at $\theta=0$. We do not impose this relation in the numerical method, but use it as a test to the code.}.
The input parameters are $r_H$, $\Omega_H$, and the parameters describing the dark matter halo, $a_0$ and $\mathcal{M}$.

To solve the system of partial differential equations, we developed a high-precision version of the code presented in Ref.~\cite{Fernandes:2022gde}, originally developed by one of the authors, which combines a pseudospectral method with the Newton-Raphson root-finding algorithm. High-precision was required to reliably obtain physical solutions, as the energy density of the dark matter halo introduces steep gradients in the metric functions. This version employs arbitrary-precision arithmetic via the \texttt{DoubleFloats.jl} library \cite{sarnoff2022doublefloats} in \texttt{Julia}. Details on code, its validation and accuracy are provided in the \hyperref[suppmaterial]{Supplemental Material}. The code is publicly available at \cite{HighPrecisionSpinningBHs}, and can be freely used for future studies of rotating BHs in galaxies.

\begin{table*}[]
    \centering
    \resizebox{\textwidth}{!}{%
        \begin{tabular}{lllllllllll}
        \toprule
        $J/M_{\rm BH}^2$ & $\delta(\kappa)$ & $\delta(A_H)$ & $\delta(R_{\rm ISCO}^{\rm co})$ & $\delta(R_{\rm ISCO}^{\rm counter})$ & $\delta(\omega_{\rm ISCO}^{\rm co})$ & $\delta(\omega_{\rm ISCO}^{\rm counter})$ & $\delta(R_{\rm LR}^{\rm co})$ & $\delta(R_{\rm LR}^{\rm counter})$ & $\delta(\omega_{\rm LR}^{\rm co})$ & $\delta(\omega_{\rm LR}^{\rm counter})$ \\
        \midrule
        0.000 & -17.2 & 20.7 & 7.5 & 7.5 & -13.4 & -13.4 & 10.0 & 10.0 & -17.1 & -17.1 \\
        0.175 & -17.0 & 21.1 & 9.9 & 5.5 & -16.2 & -10.9 & 11.3 & 8.8 & -18.2 & -16.3 \\
        0.300 & -16.6 & 21.8 & 12.0 & 4.3 & -18.6 & -9.3 & 12.4 & 8.1 & -19.2 & -15.8 \\
        0.443 & -15.8 & 23.2 & 15.2 & 3.1 & -21.8 & -7.6 & 13.7 & 7.3 & -20.7 & -15.4 \\
        0.635 & -13.1 & 26.9 & 21.3 & 1.6 & -27.6 & -5.4 & 15.7 & 6.4 & -23.6 & -14.9 \\
        0.805 & -5.6 & 34.0 & 30.9 & 0.5 & -35.8 & -3.4 & 17.4 & 5.8 & -28.0 & -14.7 \\
        0.902 & 9.2 & 43.3 & 40.9 & -0.2 & -43.9 & -2.3 & 18.2 & 5.5 & -32.7 & -14.7 \\
        0.996 & 252.6 & 78.6 & 63.2 & -0.8 & -63.3 & -1.2 & 18.7 & 5.3 & -45.0 & -14.7 \\
        % Add more rows as needed
        \bottomrule
        \end{tabular}%
    }
    \caption{Percentual deviations in physical quantities from a Kerr BH with the same $M_{\rm BH}$ and $J/M_{\rm BH}^2$, for solutions with $M_{\rm halo} \approx 10 M_{\rm BH}$, and $a_0 \approx 100 M_{\rm BH}$. All deviations are rounded to the first decimal place.}
    \label{tab:deltas}
\end{table*}

%%%%%%%%%%%%%%%%%%%%%%%%%%%%%%%%%%%%%%%%%%%%
\noindent \textbf{\textit{Results.}}
%%%%%%%%%%%%%%%%%%%%%%%%%%%%%%%%%%%%%%%%%%%%
The parameter space is vast: fixing the BH mass, it is determined by the BH spin, the mass of the halo, and the length scale $a_0$. Increasing $a_0$ while keeping the other parameters constant significantly complicates the numerical problem. For sufficiently large $a_0$ (low halo compactness $M_{\rm halo}/a_0$), the numerical method requires very high resolution due to steep gradients in the metric functions emerging within a narrow region of the numerical grid. This occurs because around the characteristic length scale, $r\sim a_0$ (large values of $r$), the energy density profile transitions between different regimes: $\varepsilon \sim 1/r$ when $r\ll a_0$, while $\varepsilon \sim 1/r^4$ when $r \gg a_0$, as is characteristic of the Hernquist profile. We discuss this issue in more detail in the \hyperref[suppmaterial]{Supplemental Material}.
For these reasons, we have focused on the case $M_{\rm halo} \approx 10 \MBH$ and $a_0 \approx 100 \MBH$, ensuring a clear hierarchy between $\MBH$, $\Mh$, and $a_0$, while keeping the computational cost manageable for a systematic study. This setup is sufficient for our primary objective: to investigate the impact of spin on BHs in astrophysical environments. We have also examined solutions with $M_{\rm halo} \approx 10 \MBH$ and $a_0 \approx 1000 \MBH$, finding that the qualitative conclusions remain consistent.

\begin{figure}[]
	\centering
	\includegraphics[width=\linewidth]{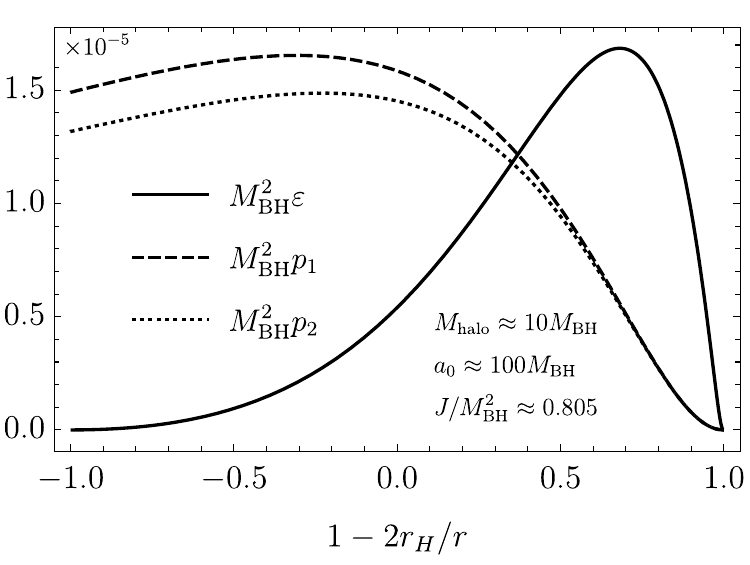}\hfill
    \caption{Profiles for the eigenvalues of the stress-energy tensor, $\varepsilon$, $p_1$ and $p_2$, for a solution with $J/M_{\rm BH}^2 \approx 0.805$, $M_{\rm halo} \approx 10 M_{\rm BH}$, and $a_0 \approx 100 M_{\rm BH}$, at the equatorial plane, $\theta=\pi/2$.}
	\label{fig:energyconditions}
\end{figure}

First, we examined whether rotation induces violations of the standard energy conditions by studying the eigenvalues of the stress-energy tensor. The energy conditions are as follows. Null energy condition: $\varepsilon + p_i \geq 0$; Weak energy condition: $\varepsilon\geq 0$ and $\varepsilon + p_i \geq 0$; Strong energy condition: $\varepsilon + p_i \geq 0$ and $\varepsilon + \sum_i p_i \geq 0$;  Dominant energy condition: $\varepsilon \geq |p_i|$. Here, $i$ runs over $\{r, 1, 2\}$.
In the static case originally studied in Ref. \cite{Cardoso:2021wlq}, it was found that only the dominant energy condition is violated, in a small, almost empty, region close to the horizon. However, simple
modifications of the original profile, e.g. by using a cutoff radius in the profile for the energy density, eliminate this issue \cite{Cardoso:2021wlq,Speeney:2024mas,Shen:2024qbb,Shen:2023erj}. Since, for simplicity, we do not impose such a cutoff radius in our profile for the energy density, Eq. \eqref{eq:edensity}, we expect similarly to Ref. \cite{Cardoso:2021wlq}, violations of the dominant energy condition in a small region close to the horizon. However, these modifications of the energy-density profile are not expected to significantly affect the physical solution, since they occur in an almost empty region with negligible energy density and therefore have no influence on the spacetime dynamics.
In Fig. \ref{fig:energyconditions}, we show the equatorial plane profiles for the eigenvalues of the stress-energy tensor for a solution with $J/\MBH^2 \approx 0.8$. All eigenvalues remain non-negative throughout the domain, indicating that the null, weak, and strong energy conditions are satisfied. As anticipated, the dominant energy condition is violated near the horizon. Similar profiles are observed for other values of $\theta$, and the same conclusions apply to solutions with different spins. Thus, once a suitable modified profile for the energy density is used, we expect that solutions in this framework to obey all energy conditions.

To analyze the impact of rotation, we have calculated various physical quantities of BHs surrounded by a dark matter halo and compared them to those of a Kerr BH with the same mass $\MBH$ and dimensionless spin $J/\MBH^2$. To quantify deviations from the Kerr metric, we define
\begin{equation}
    \delta(x) = \left(\frac{x}{x_{\rm Kerr}}-1\right)\times 100,
    \label{eq:delta}
\end{equation}
where $x$ is any physical quantity.
The quantities considered include the surface gravity and event horizon area (both defined in Eq. \eqref{eq:kappaAH}), as well as the locations and frequencies of circular orbits for massive and massless particles at the innermost stable circular orbit (ISCO) and at the light ring (LR), respectively, for both co-rotating and counter-rotating motions. Ref. \cite{Fernandes:2022gde} explains in detail how to compute these quantities for a line-element of the form \eqref{eq:metric}.
These quantities are important because, for instance, the location of the ISCO is commonly regarded as the inner edge of the accretion disk around a BH. Additionally, the characteristics of synchrotron radiation emitted by charged particles orbiting near the BH are linked to the orbital frequencies at the ISCO. The LR is closely connected to the BH shadow, and the geodesic frequencies at the LR relate to the characteristic timescale of the BH response to perturbations.
The location of the circular orbits is measured using the geometrically meaningful perimetral radius, $R = \sqrt{g_{\varphi \varphi}}|_{\theta=\pi/2}$.

The deviations are presented in Tab.~\ref{tab:deltas}. We observe that deviations from the Kerr metric increase with spin for the event horizon area $A_H$ and for quantities related to co-rotating orbits, while they decrease with spin, up to a point, for the surface gravity $\kappa$ and for quantities associated with counter-rotating orbits. The location of the ISCO and LR of galactic BHs lie, in general, at greater radii compared to a vacuum Kerr BH, and the corresponding geodesic frequencies are consequently smaller. While the extremal limit for a Kerr BH is reached for $J/\MBH^2 = 1$, we find that for galactic BHs this limit occurs at higher values of spin. We were able to find configurations with $J/\MBH^2 > 1$, and therefore it might be possible for astrophysical black holes to surpass the Thorne limit \cite{1974ApJ...191..507T}.
For configurations with $\Mh/a_0 \approx 10^{-2}$, we observed that deviations, in general, decreased one order of magnitude compared to those of Tab.~\ref{tab:deltas}.

Following the first images of M87* \cite{EventHorizonTelescope:2019dse,EventHorizonTelescope:2019uob,EventHorizonTelescope:2019jan,EventHorizonTelescope:2019ths,EventHorizonTelescope:2019pgp,EventHorizonTelescope:2019ggy,EventHorizonTelescope:2021bee,EventHorizonTelescope:2021dqv,EventHorizonTelescope:2021srq,EventHorizonTelescope:2023gtd}, Sgr A* \cite{EventHorizonTelescope:2022wkp,EventHorizonTelescope:2022apq,EventHorizonTelescope:2022wok,EventHorizonTelescope:2022exc,EventHorizonTelescope:2022urf,EventHorizonTelescope:2022xqj,EventHorizonTelescope:2024hpu,EventHorizonTelescope:2024rju} and their subsequent observations \cite{EventHorizonTelescope:2024dhe,EventHorizonTelescope:2024uoo}, considerable attention has been devoted to BH imaging in recent years, see Refs. \cite{Cunha:2018acu,Lupsasca:2024wkp} (and references therein) for reviews.
When light rays are traced backwards from the observer, geodesics fall into two distinct classes: those that eventually plunge into the BH and those that escape to infinity \cite{Cunha:2018acu}. The boundary between these two classes is known as the \emph{critical curve} (or sometimes, shadow), and is fully determined by the spacetime geometry.
Surrounding the critical curve is a bright region known as the \emph{photon ring} \cite{Johnson:2019ljv,Gralla:2019xty,Gralla:2020srx,Gralla:2020yvo,Broderick:2021ohx,Chael:2021rjo,Lupsasca:2024wkp}, composed of an infinite sequence of stacked subrings. The $n^{\rm th}$ subring consists of photons that perform $n$ half-orbits around the BH before escaping. These subrings converge exponentially to the critical curve, and the $n^{\rm th}$ ring always lies within the $n^{\rm th}$ \emph{lensing band} \cite{Cardenas-Avendano:2023obg}: a region of the observer's screen consisting of all possible light rays that complete at least $n$ half-orbits around the BH along their trajectories.

To explore image features of galactic BHs, and how they change with spin, we backwards ray-traced null geodesics in the galactic BH background metrics obtained through our numerical method. This was done using the ``Flexible Object Oriented Ray Tracer'' (\texttt{FOORT}) code \cite{Mayerson:2025xxx}, which has previously been used to analyze image features of compact objects in modified theories of gravity \cite{Staelens:2023jgr,Mayerson:2023wck,Carballo-Rubio:2025zwz,Fernandes:2024ztk}. An observer is placed at a distance $r = 1000 \MBH$ from the BH, with a field of view of $20\MBH \times 20\MBH$, and at some inclination. Geodesics are traced until they either reach the event horizon, or a celestial sphere also located at $r = 1000\MBH$, while we keep track the number of half-orbits each geodesic completes around the BH, to compute the respective lensing band. Motivated by the M87* BH, we adopted $17^\circ$ as a reference inclination, and used a resolution of $1024\times 1024$ pixels.

\begin{table}[]
    \centering
    %\resizebox{\linewidth}{!}{%
        \begin{tabular}{ll|ll}
        \toprule
        $J/M_{\rm BH}^2$ & $\delta(A_{\rm sh})$ & $J/M_{\rm BH}^2$ & $\delta(A_{\rm sh})$ \\
        \midrule
        0.000 & 40.6 & 0.635 & 44.0 \\
        0.175 & 40.8 & 0.805 & 46.6 \\
        0.300 & 41.3 & 0.902 & 49.0 \\
        0.443 & 42.1 & 0.996 & 52.5 \\
        % Add more rows as needed
        \bottomrule
        \end{tabular}%
    %}
    \caption{Percentual deviations in the shadow area relative to a Kerr BH with the same $M_{\rm BH}$ and $J/M_{\rm BH}^2$ are shown for solutions with $M_{\rm halo} \approx 10 M_{\rm BH}$ and $a_0 \approx 100 M_{\rm BH}$.}
    \label{tab:shadows}
\end{table}
Although the critical curve itself is \emph{not} an observable, since astrophysical emission may or may not produce a brightness deficit that aligns with it, the detection of a sufficiently high-$n$ photon ring can yield an accurate estimate of its diameter. Since our main interest is in how spin influences physical quantities, we have investigated the critical curve of galactic BHs. Specifically, we computed the area of the critical curve/shadow, $A_{\rm sh}$, and compared it to that of Kerr BHs with the same mass and spin, via $\delta(A_{\rm sh})$. The results are presented in Tab. \ref{tab:shadows}. We observe that the area of the critical curve of galactic BHs is greater than that of a Kerr BH with the same mass and spin, in agreement with the results of Ref. \cite{Xavier:2023exm} in the static case, and that these deviations increase with spin.

\begin{figure}[]
	\centering
	\includegraphics[width=\linewidth]{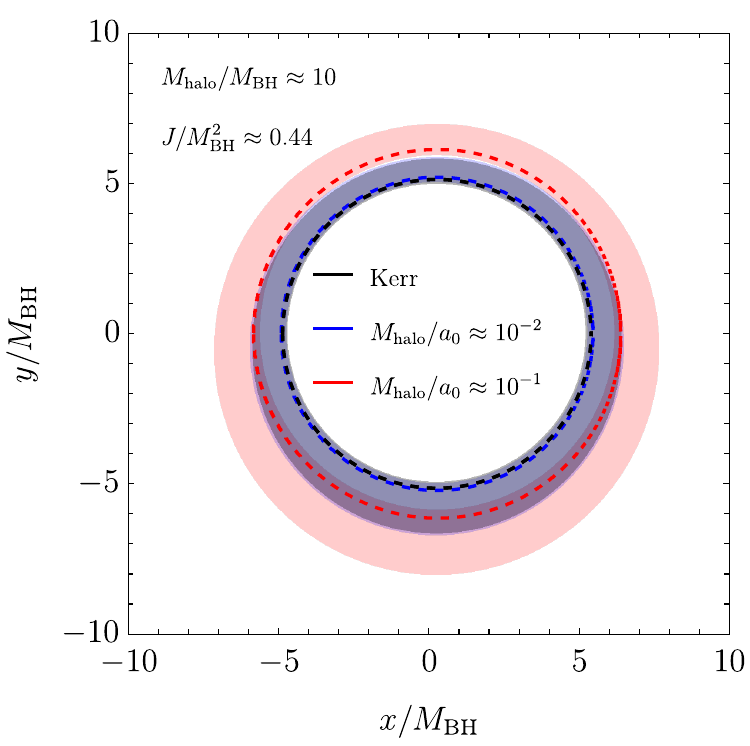}\hfill
    \caption{Critical curves (dashed lines), and $n=1$ lensing bands (colored regions) for a Kerr BH (gray) and for galactic BHs with $\Mh \approx 10\MBH$, for compactnessess $\Mh/a_0 \approx 10^{-1}$ (red) and $\Mh/a_0 \approx 10^{-2}$ (blue). The spin of the BHs is $J/\MBH^2 \approx 0.44$. The darker colored regions correspond to the intersection between the different lensing bands.}
	\label{fig:shadows}
\end{figure}
The large deviations from the Kerr metric observed in Tab. \ref{tab:shadows} are, however, suppressed by the compactness of the halo.
For less compact halos, deviations become small across all spin values. As an example, in Fig. \ref{fig:shadows}, we show the critical curves and the $n = 1$ lensing bands for a Kerr BH and for galactic BHs with $\Mh \approx 10\MBH$, considering compactness ratios $\Mh/a_0 \approx 10^{-1}$ and $\Mh/a_0 \approx 10^{-2}$. We find that for the low-compactness halo with $\Mh/a_0 \approx 10^{-2}$, deviations in both the critical curve and the $n = 1$ lensing band are barely perceptible, at the percent level, in contrast to the case with $\Mh/a_0 \approx 10^{-1}$. This suggests that tests of the nature of central galactic objects can be performed, e.g. with the Event Horizon Telescope, to good precision in realistic settings.

%%%%%%%%%%%%%%%%%%%%%%%%%%%%%%%%%%%%%%%%%%%%
\noindent \textbf{\textit{Discussion.}}
%%%%%%%%%%%%%%%%%%%%%%%%%%%%%%%%%%%%%%%%%%%%
In this letter, we have introduced a formalism to describe an anisotropic fluid in a rotating BH spacetime and used it to construct solutions to the Einstein equations representing a rotating BH in a galactic environment. Our work marks a crucial step forward in the study of BHs in astrophysical environments by incorporating rotation, a feature expected to be generic for astrophysical BHs and essential for understanding their physical and observational properties.

We have found that deviations from the Kerr metric in physical quantities generally increase with spin. Interestingly, we found that galactic BHs can have spins $J/\MBH^2 >1$, and thus have the potential to surpass the Thorne limit. The energy conditions follow a pattern similar to the static case, with only the dominant energy condition violated in a small region near the horizon. However, this issue can likely be resolved through a simple modification of the energy density profile, for example, by truncating it below a certain threshold radius. This suggests that the Einstein cluster remains a realistic model for the galactic environment surrounding a rotating BH.

Although we adopted a specific fluid rotation law, $\Omega=\omega$, to remain consistent with the Einstein cluster construction, alternative rotation laws, where the surrounding environment also carries angular momentum, are valid and could be explored in future work.
Within our framework, and for realistic configurations, we find no evidence for abnormal rotation curves: timelike geodesics obey Kepler's law to a good approximation, dictated by the mass enclosed within the orbit. This is in agreement with the general result for Newtonian-like distributions of matter~\cite{Duque:2023seg} and hence one cannot explain dark matter as a general relativistic effect~\cite{Costa:2023awm}.

The code used to generate the rotating BH solutions is publicly available \cite{HighPrecisionSpinningBHs} and can be freely used for future studies. In this context, extending many of the analyses performed in the static case to the rotating case -- particularly those related to gravitational wave phenomena -- would be especially valuable. An additional direction for future research, specific to the rotating case, involves investigating the non-integrability of geodesics and the emergence of chaos due to the environment, along with their imprints on gravitational waves~\cite{Apostolatos:2009vu,Destounis:2023khj,Destounis:2021rko,Destounis:2021mqv,Destounis:2020kss}.

%%%%%%%%%%%%%%%%%%%%%%%%%%%%%%%%%%%%
\noindent {\bf \em Acknowledgments.} 
%%%%%%%%%%%%%%%%%%%%%%%%%%%%%%%%%%%
We thank Kyriakos Destounis, Andrea Maselli, José Natário, and Prashant Kocherlakota for useful comments on a previous version of the manuscript.
P.F. would like to thank CENTRA, Instituto Superior Técnico, for their kind hospitality during the initial stages of this work.
The Center of Gravity is a Center of Excellence funded by the Danish National Research Foundation under grant No. DNRF184.
We acknowledge support by VILLUM Foundation (grant no. VIL37766) and the DNRF Chair program (grant no. DNRF162) by the Danish National Research Foundation.
V.C.\ is a Villum Investigator and a DNRF Chair.  
V.C. acknowledges financial support provided under the European Union’s H2020 ERC Advanced Grant “Black holes: gravitational engines of discovery” grant agreement no. Gravitas–101052587. 
Views and opinions expressed are however those of the author only and do not necessarily reflect those of the European Union or the European Research Council. Neither the European Union nor the granting authority can be held responsible for them.
This project has received funding from the European Union's Horizon 2020 research and innovation programme under the Marie Sklodowska-Curie grant agreement No 101007855 and No 101131233. 
P.F. is funded by the Deutsche Forschungsgemeinschaft (DFG, German Research Foundation) under Germany’s Excellence Strategy EXC 2181/1 - 390900948 (the Heidelberg STRUCTURES Excellence Cluster).

\bibliography{References}

\onecolumngrid
\newpage
\setcounter{equation}{0}
\renewcommand{\theequation}{\roman{equation}}
\section*{Supplemental Material}
\label{suppmaterial}

\section*{Static and spherically symmetric analytic solution, and the energy density profile}

In Schwarzschild-type coordinates there are known analytic solutions to the Einstein equations sourced by an Einstein cluster, in the absence of rotation. However, Schwarzschild-type coordinates are not well-suited for a numerical approach once rotation is considered, which is the reason we employ quasi-isotropic coordinates in our code. Therefore, we cannot directly use the energy density profile obtained in Ref. \cite{Cardoso:2021wlq}. Moreover, for the solution of Ref. \cite{Cardoso:2021wlq}, the coordinate transformation to isotropic coordinates is not integrable analytically.
Similarly to Ref. \cite{Zhang:2024hjr}, we have tried to perform an \emph{approximate} coordinate transformation on the energy density profile, but found this approach to be ill suited, as the energy density and its parameters lose their physical meaning. 

To solve this issue, in this section we try to solve, at least partially, the Einstein equations in the absence of rotation and in isotropic coordinates to find a suitable profile for the energy density. In this case, the metric functions $f$ and $g$ depend only on $r$, while $h=1$, $\omega=0$. Defining $b(r)^4 = f N_+^4/g$, the $tt$ component of the Einstein field equations becomes
\begin{equation}
    \frac{8 b' + 4 r b''}{r b^5} = 8\pi \varepsilon,
    \label{eq:diffeqstatic}
\end{equation}
where a prime denotes a radial derivative. Motivated by the Hernquist profile, by Ref. \cite{Cardoso:2021wlq}, and by the form of this differential equation, we consider the following ansatz for the energy density
\begin{equation}
    \varepsilon = \frac{\mathcal{M} \left( a_0 + r_H \right)}{2\pi r (r+a_0)^3} \left(1 - \frac{r_H}{r} \right)^2 b^{-5},
    \label{eq:edensity_supp}
\end{equation}
where $\mathcal{M}$ and $a_0$ are parameters to be determined based on physical considerations. In this case, the differential equation \eqref{eq:diffeqstatic} can be solved for $b$ analytically, yielding
\begin{equation}
    \begin{aligned}
        &b(r) = \left(1 + \frac{r_H}{r}\right) \bigg[ 1 + \frac{\mathcal{M} \left( a_0 + r_H \right)}{2a_0^4(r+r_H)(r+a_0)} \bigg( a_0  \left(r \left(a_0 ^2+12 r_H^2+3 a_0  r_H\right)+a_0  r_H (a_0 +11 r_H)\right) \\& +2 r_H (a_0 +r) \left(r (2 a_0 +3 r_H) \log \left(\frac{r}{a_0 +r}\right)+r_H \left(a_0  \log \left(\frac{r_H r}{(a_0 +r) (a_0 +r_H)}\right)+3 r_H \log \left(\frac{r_H}{a_0 +r_H}\right)\right)\right) \bigg) \bigg],
    \end{aligned}
    \label{eq:analytic_sol_supp}
\end{equation}
once suitable boundary conditions are imposed.

The asymptotic decay of $g_{rr}$ follows
\begin{equation}
    g_{rr} \approx 1 + \frac{2M}{r},
\end{equation}
from which, using the analytic solution, we determine that to leading order in $\mathcal{M}/a_0$ we have $\mathcal{M} = M_{\rm halo}$, and $a_0$ represents a length scale associated with the dark matter halo. Importantly, $b>0$, and thus $\varepsilon \geq 0$ for physically relevant choices of the parameters $\mathcal{M}$ and $a_0$.
In the large $r$ limit, this profile for $\varepsilon$ agrees with the Hernquist profile, up to terms proportional to $r_H$.

\section*{The code}
The code we developed is a high-precision version of the code presented in Ref.~\cite{Fernandes:2022gde}. It uses a pseudospectral method and the Newton-Raphson root-finding algorithm to solve the partial differential equations. This version employs arbitrary-precision arithmetic via the \texttt{DoubleFloats.jl} library \cite{sarnoff2022doublefloats} in \texttt{Julia}. For the pseudospectral method, we use the compactified radial coordinate $x = 1-\frac{2r_H}{r}$, which maps the radial domain to the finite range $x \in [-1,1]$.
Each of the functions is expanded in a spectral series with resolution $N_x$ and $N_\theta$ in the radial and angular coordinates $x$ and $\theta$, respectively. The spectral series used for each of the functions $\mathcal{F}^{(k)}=\{f,g,h,\omega,p_1,p_2\}$ is
\begin{equation}
  \mathcal{F}^{(k)} = \sum_{n=0}^{N_x-1} \sum_{m=0}^{N_\theta-1} c_{nm}^{(k)} T_n(x) \cos \left(2m\theta\right),
\label{eq:spectralexpansion1}
\end{equation}
where $T_n(x)$ denotes the $n^{\rm th}$ Chebyshev polynomial, and $c_{nm}^{(k)}$ are the spectral coefficients. The angular boundary conditions are automatically satisfied with this spectral expansion, and need not be explicitly imposed in the numerical method. Convergence was declared once the norm of the difference between the spectral coefficients of successive iterations was smaller than a prescribed tolerance, fixed at $10^{-10}$.

We successfully reproduced the Kerr metric with precision beyond machine precision, obtaining errors smaller than $10^{-16}$ in all physical quantities. Furthermore, we observed exponential convergence as the numerical resolution was increased.

To further validate the convergence of our code, we employed several independent approaches. First, in the static limit, we compared the numerical solution with the analytic expression given in Eq. \eqref{eq:analytic_sol_supp}, finding excellent agreement and convergence as the resolution was increased. Second, we verified that the Komar and ADM masses converge to the same value as the resolution increases. Third, we checked that the surface gravity at the horizon approaches a constant value with increasing resolution.

\begin{figure}[]
	\centering
	\includegraphics[width=0.5\linewidth]{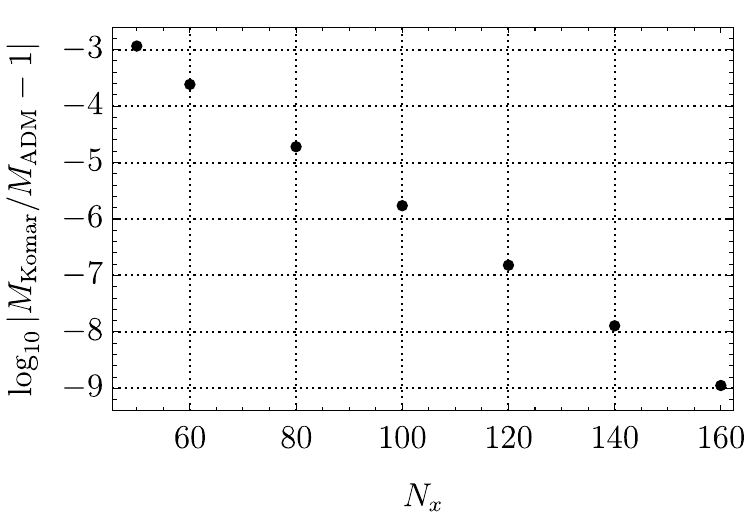}\hfill
	\includegraphics[width=0.5\linewidth]{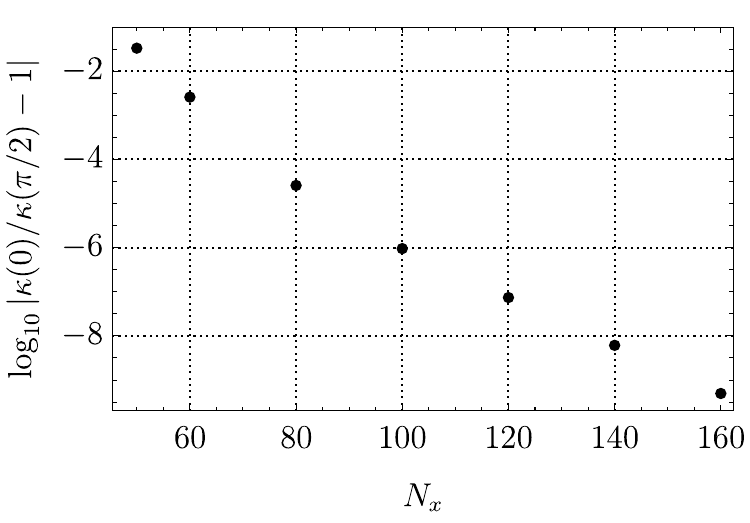}\vfill
    \caption{Convergence tests for a solution with $J/\MBH^2 \approx 0.635$, $\Mh\approx 10\MBH$, and $a_0 \approx 100 \MBH$. The resolution in the angular coordinate is fixed as $N_\theta=4$.}
	\label{fig:convergence}
\end{figure}

In Fig. \ref{fig:convergence}, we present convergence tests for a solution with $J/\MBH^2 \approx 0.635$, $\Mh \approx 10 \MBH$, and $a_0 \approx 100 \MBH$, performed by fixing $N_\theta$ and varying $N_x$. The left panel shows the deviation between the Komar and ADM masses, while the right panel shows the difference in surface gravity computed at $\theta = 0$ and $\theta = \pi/2$, as functions of $N_x$. In both cases, we observe clear exponential convergence as $N_x$ increases. Similar convergence tests were performed for other solutions, and we typically used resolutions such that the maximum estimated error was of order $10^{-8}$ or lower.

\section*{Numerical issues for large characteristic length scale $a_0$}
The scale $a_0$ marks the characteristic scale at which the energy density of the halo transitions from $\varepsilon \sim 1/r$ when $r\ll a_0$, to $\varepsilon \sim 1/r^4$ when $r \gg a_0$. This leads to steep gradients in the metric functions, which are difficult to capture with the numerical method.

\begin{figure}[]
	\centering
	\includegraphics[width=0.5\linewidth]{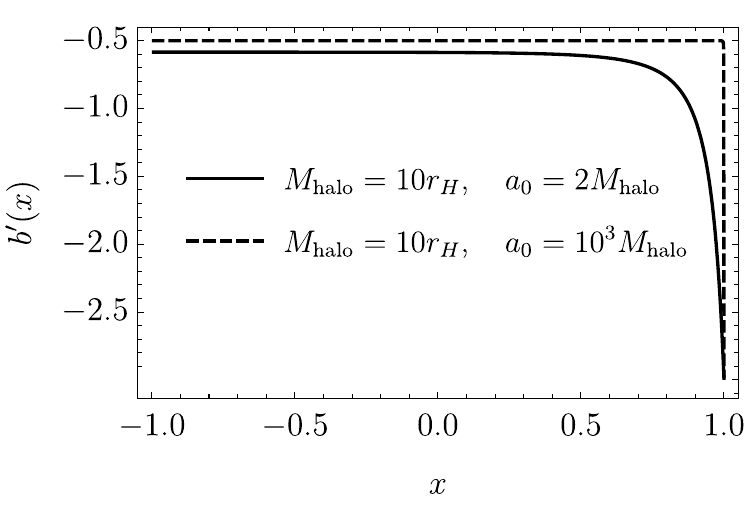}\hfill
    \caption{$b'(x)$ as a function of $x=1-2r_H/r$, for $M_{\rm halo} = 10r_H$, and two distinct values of $a_0$. We observe steep gradients when $a_0$ is large. For comparison, in the vacuum GR case, $b(x)=(3-x)/2$, such that $b'(x)=-1/2$.}
	\label{fig:gradients}
\end{figure}

As an example, we consider the static limit, where we know the metric function $b(r)^4 = f N_+^4/g$ analytically, as obtained in the sections above. In Fig. \ref{fig:gradients} we plot $b'(x)$ as a function of $x=1-2r_H/r$, which is the radial coordinate employed numerically. We observe that when $a_0$ is large, there are steep gradients in the metric functions, which appear in a small range of $x$, that a numerical method struggles to capture accurately. For this reason, for large values of $a_0$, a (very) large resolution in $x$ must be used.

\end{document}